\documentclass[journal]{IEEEtran}

  \ifCLASSINFOpdf

  \else

  \fi

\usepackage{multirow}

  \usepackage{amsfonts}
  \usepackage{graphicx}
  \graphicspath{ {images/} }
  \usepackage{caption}
  \usepackage{subcaption}
  \usepackage{cite}
  \usepackage{graphicx}
  \usepackage{psfrag}
  \usepackage{url}
  \usepackage{stfloats}
  \usepackage{amsmath}
  \usepackage{array}
  \usepackage{amssymb}
  \usepackage{setspace}
  \usepackage{algorithmicx}
  \usepackage[ruled]{algorithm}
  \usepackage{algpseudocode}
  \usepackage{algpascal}
  \usepackage{algc}
  \usepackage{color}

\usepackage{url}
\usepackage{lipsum}
\usepackage{blkarray}
\usepackage{xcolor}
\usepackage{soul}
\usepackage{booktabs}
\usepackage{multirow}
\usepackage{microtype}
\usepackage{epstopdf}

\begin{document}

  \title{\LARGE Distortion-free Golden-Hadamard Codebook Design for MISO Systems}

  \author{Md. Abdul Latif Sarker, Md. Fazlul Kader, Moon Ho Lee, and Dong Seog Han\vspace{-4ex}
  \thanks{Manuscript received July 5, 2018; revised July 26, 2018; accepted July 29, 2018. This work was supported by the Brain Korea 21 (BK21) Plus project funded by the Korean Ministry of Education (21A20131600011) and a Korea Institute for Advancement of Technology (KAIT) grant funded by the Korean Government (No.P0000535). The associate editor coordinating the review of this paper and approving it for publication was A. Garcia Armada. (\textit{corresponding author: Dong Seog Han.})}
  \thanks {M. A. L. Sarker and D. S. Han are with the School of Electronics Engineering, Kyungpook National University, Daegu 41566, South Korea. (e-mail: latifsarker@jbnu.ac.kr; dshan@knu.ac.kr).}
  \thanks {M. F. Kader is with the Department of Electrical and Electronic Engineering, University of Chittagong, Chittagong 4331, Bangladesh (email: f.kader@cu.ac.bd).}
  \thanks{M. H. Lee is with the Department of Electronics Engineering, Chonbuk National University, Jeonju 54896, South Korea. (e-mail: moonho@jbnu.ac.kr).}
  \thanks { Digital Object Identifier: 10.1109/LCOMM.2018.2864124}}

  \markboth{IEEE Communication Letters,~Vol.~xx, No.~xx, xx~2018}%
  {}


  \maketitle

  \begin{abstract}
    In this letter, a novel Golden-Hadamard codebook (GHC) scheme is proposed to improve the performance of the traditional precoded Alamouti coding for multiple-input and single-output systems. Although the traditional discrete Fourier transform codebook (DFTC) performs satisfactorily with Alamouti coding and offers numerous benefits for the Rayleigh fading channel, this scheme inherently generates huge codeword distortion, which leads to a lower minimum chordal distance (MCD). Furthermore, the uncertain format of all prior versions of codebooks results in poorer minimum determinant (MD) values. Hence, the proposed GHC scheme successfully deals with the issues of traditional DFTC to achieve a better codebook format that completely overcome both MCD and MD problems. The effectiveness of the proposed GHC scheme is confirmed, in terms of bit-error-rate through Monte Carlo simulations.
  \end{abstract}
  \IEEEpeerreviewmaketitle
  \section{Introduction}
 \IEEEPARstart{S}{pace-time} block coding (STBC) is an important technique used in wireless communications to improve diversity gain \cite{1} and data rates \cite{2}. Orthogonal STBC (OSTBC) allows low-complexity decoding \cite{3} and does not provide array gain. Thus, precoded OSTBC (POSTBC), such as the discrete Fourier transform codebook with Alamouti (DFTC-A) coding, has been proposed \cite{4} to improve array gain and attain low average codeword distortion using a matrix search method. In contrast, in another study, a vector search method with unitary and non-unitary POSTBC \cite{5} to minimize codebook selection complexity. Additionally, Suh et al. proposed using the DFTC \cite{6} to achieve high eigenvectors with a high resolution. In \cite{7}, Hai et al. compared the performance of the discrete fractional sine transforms codebook (DFRSTC) with that of the DFTC, demonstrating that they are equal. The Hadamard codebook (HC) \cite{8} and the diagonal codebook (DC) \cite{9} has been proposed along with OSTBC to improve error performance. Furthermore, Grag et al. showed their use of a DC in a non-OSTBC scheme \cite{10}. Unfortunately, the uncertain format of all previous version of POSTBC schemes produce lower minimum chordal distances (MCDs) and minimum determinant (MD) values. Based on this problem, we propose a Golden-Hadamard (GH) codebook with Alamouti coding scheme (GHC-A) in this letter. The Golden code has been proposed along with STBC \cite{11}. Although the Golden code has a higher decoding complexity \cite{11}, such schemes show a better bit-error-rate (BER) performance \cite{10}. Based on the literature \cite{4} and \cite{9, 10, 11}, a newly design GHC generator can lead an identical BER performance.\par
 In this letter, we present a novel GHC-A scheme to achieve a higher MCD and MD values, which are key factors to demonstrate good performance for any precoded codeword. Developing a better codebook generator is the principal subject of this letter. However, a limited perfect and imperfect feedback scheme is considered in computer simulations. The superiority of the proposed GHC-A scheme over the traditional POSTBC schemes for multiple-input single-output (MISO) systems, in terms of BER is verified through computer simulations.
 \section{System Model}
We consider a MISO system where the transceiver is equipped with $N_{T}$ transmit antennas  and single receive antenna. Let $N_{T}$ channels remains static over the symbol period, $T$. Then the received signal $\mathbf{y} \epsilon {\mathbb{C}}^{1\times T}$  can be modeled as
 \begin{equation}
 \textbf{y}=\textbf{h}^{T}\textbf{X}+\textbf{z}
 \end{equation}
 where $\textbf{h}$ is the flat and block fading channel vector between the transmitter and receiver, $(\cdot)^{T}$ represents transpose operation, and  a high-order $N_{T}\times T$  precoded codeword, $\textbf{X}$ is
 \begin{equation}
 \mathbf{X}=\mathbf{W_{D}}\mathbf{S}
 \end{equation}
 where $\mathbf{S}$ is a low-order $M\times T$ OSTBC, $M\leq N_{T}$, $\mathbf{W_{D}}$ is an $N_{T}\times M$ tall DFT precoding matrix with $\frac{1}{\sqrt{N_{T}}} e^{\frac{\textit{j}2\pi kl}{N_{T}}}$ at entry $(k,l)$ which is chosen from the codebook generator presented in \cite{4,5,7}:
 \begin{equation}
 {\mathcal{F}}_{D}=\{{\mathbf{\Theta}}_{D}^{\textit{i}-1}\mathbf{W}_{D}\},\textit{i}=1,2,...,L-1
 \end{equation}
 The ${N_{T}\times M}$ remaining DFT precoding matrices are given by
 \begin{equation}
 \mathbf{W}_{D,\textit{i}}=\mathbf{\Theta}_{D}^{\textit{i}-1}\mathbf{W}_{D,1}, \textit{i}=2,...,L
 \end{equation}
 where $L$ is the size of a codebook and $\mathbf{W}_{D,1}$ is the first $N_{T}\times M$ DFT-based precoding matrix, of which entry $(k,l)$  is given as $\frac{1}{\sqrt{N_{T}}} e^{\frac{j2\pi (k-1)(l-1)}{N_{T}}}$, $\mathbf{\Theta}_{D}$ is a diagonal matrix \cite{4}, and $\textbf{z} \epsilon  {\mathbb{C}}^{1\times T}$ is a noise vector with zero mean and unit variance.
\section{The Problem of the Existing DFTC Schemes}
The given DFTC entails two main problems in the POSTBC scheme. The first problem is codeword distortion which leads to a lower MCD value, and the second problem is the uncertain format of DFTC causing a smaller MD value.\\
\textit{A. \textbf{Codeword distortion}}: When a precoded codeword block or blocks contain non-zero diagonal elements, this is denoted as codeword distortion or geometric mean distortion. A geometric mean is a special kind of mean value calculated by multiplying elements and taking their square root (for two elements), their cube root (for three elements), and so on. Let the square root of the product of two matrices as
 $\sqrt{\left(\begin{IEEEeqnarraybox*}[][c]{c/c/c,c/c/c}
  2&.08\\
  .08&2
  \end{IEEEeqnarraybox*}\right)
  \left(\begin{IEEEeqnarraybox*}[][c]{c/c/c,c/c/c}
  .08&2\\
  2&.08
  \end{IEEEeqnarraybox*}\right)}
  =\left(\begin{IEEEeqnarraybox*}[][c]{c/c/c,c/c/c}
  \sqrt{.32}&2\\
  2&\sqrt{.32}
  \end{IEEEeqnarraybox*}\right)$,
  where diagonal element $\sqrt{.32}\neq 0$ which represents a non-zero geometric mean, is directly related to an increase in codeword distortion.\\
  \emph{\textbf{Example 1}:} Let a $2\times 2$ DFTC-A third codeword matrix be given using by (2-4),\cite{4} as follows:
  \begin{eqnarray}
  \left [\mathbf{X}_{3}\right]& =&\frac{\mathbf{\Theta}_{D}^{2}}{\sqrt{2}} \left[\begin{IEEEeqnarraybox*}[][c]{,c/c/c,}
  1&1\\
  1&e^{\textit{j}\pi}
  \end{IEEEeqnarraybox*}\right]
  \left[\begin{IEEEeqnarraybox*}[][c]{,c/c/c,}
  s_{11}&-s_{21}^{*}\\
  s_{21}&s_{11}^{*}
  \end{IEEEeqnarraybox*}\right] \nonumber\\
  &=&\left[\begin{IEEEeqnarraybox*}[][c]{,c/c/c,}
  0.0000+0.0000\textit{i}&0.8315-0.5556\textit{i}\\
  0.1379+0.6935\textit{i}&0.1379+0.6935\textit{i}
  \end{IEEEeqnarraybox*}\right]
  \end{eqnarray}
  where $\textit{i}=3$, ${N_{T}}=2$, ${M}=2$, $\mathbf{W}_{D,\textit{3}}=\mathbf{\Theta}_{D}^{2}\mathbf{W}_{D,1}$, the rotation vector $\mathbf{u}$=[1 7] uses the IEEE 802.16e parameters, $\mathbf{\Theta}_{D}=\textrm{diag}(\left[e^{j\pi/2},e^{j7\pi/2}\right])$, $\mathbf{W}_{D,1}$ is the $2\times 2$  DFT first precoding matrix,
 $\mathbf{S}_{k}$ is the $k$-th entry of the $2\times 2$ complex Alamouti code, symbols
  $s_{11}$ and $s_{21}$ belong to a quadrature amplitude modulation (QAM) constellation, and $(\cdot)^{*}$ is the complex conjugate operator.
  Now, we increase the size of the DFT precoding in (5). Let $\textit{i}=3$, ${M}=2$, and ${N_{T}}=4$ in (5); and we obtain the $4\times 2$ DFTC-A third codeword as follows.
  \begin{eqnarray}
  \left [\mathbf{X}_{3}\right]=\left[\begin{IEEEeqnarraybox*}[][c]{,c/c/c,}
  0.0000+0.0000\textit{i}&0.8315-0.5556\textit{i}\\
  0.1379+0.6935\textit{i}&0.1379+0.6935\textit{i}\\
  0.0000-1.0000\textit{i}&-0.0000-0.0000\textit{i}\\
  0.7071-0.0000\textit{i}&-0.7071+0.0000\textit{i}
  \end{IEEEeqnarraybox*}\right].
  \end{eqnarray}
  Using (5) and (6), we plotted in Fig. 1. Fig. 1(a) depicts a low codeword distortion curve at low-order DFTC, while Fig. 1(b) shows how distortion gradually increases for high-order DFTCs. This problem dramatically lowers the MCD value. Thus, MCD is the major issue for subspace packing. The selected precoding matrix $\mathbf{W}_{D,i}^{sel}$ can be treated as the distortion due to non-ideal precoding $\mathbf{W}_{D,i}$. Therefore, a packing can be described by its MCD \cite{4, 5, 7} as following:
  \begin{eqnarray}
  \delta_{\textup{min}}(\textbf{W}_{D})=\min_{2\leq i\leq L}d(\mathbf{W}_{D,1},\mathbf{W}_{D,i})
  \end{eqnarray}
  where $d(\cdot,\cdot)$ is called the chordal distance between two subspaces $\mathcal{P}_{\textbf{W}_{D,1}}$ and $\mathcal{P}_{{\textbf{W}_{D,i}}}$, of which $\mathcal{P_{\textbf{W}_{D}}}$ is the subspace generated by the columns of matrix $\textbf{W}_{D}$.\\
 \textit{B. \textbf{MD for Any Codeword}}: The MD is an important key factor to explain the good performance of any codeword. Thus, the MD of finite code $\mathcal{C}$ is given by \cite{11}
 \begin{equation}
 \delta_{\textup{min}}(\mathcal{C})\geq  2^{b}\delta_{\textup{min}}(\mathcal{C_{\infty}})
 \end{equation}
 where $\mathcal{C_{\infty}}$ is denoted as the infinite code, $b$ is bits per symbol and  $\delta_{\textup{min}}(\mathcal{C}_{\infty})=min_{X\epsilon \mathcal{C}_{\infty},X\neq0}|det(\mathbf{X})|^{2}$.

 \begin{figure}
  \begin{minipage}[b]{.51\linewidth}
  \centering\includegraphics[width=45mm,height=27mm]{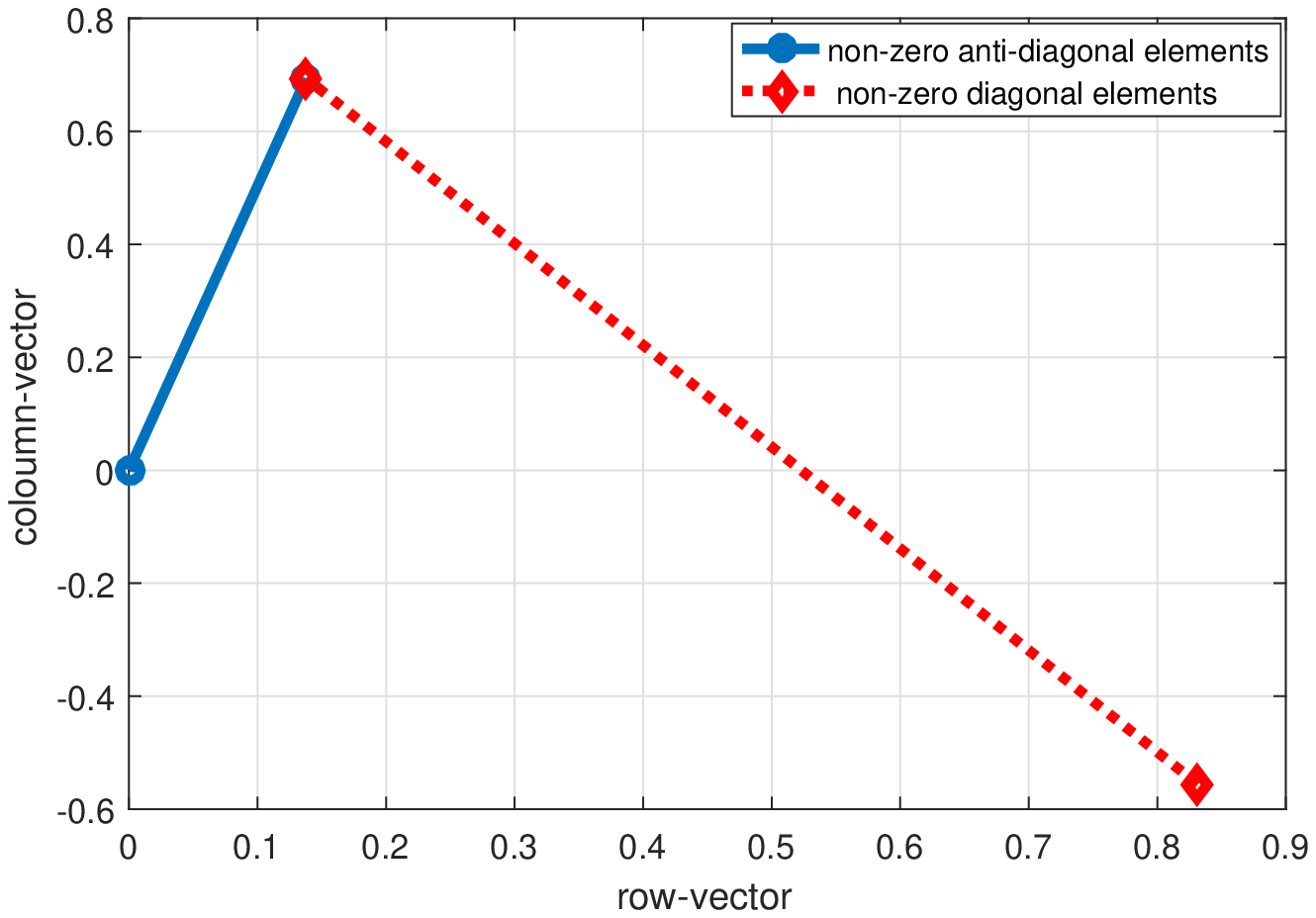}
  \subcaption{}
  \end{minipage}
  \begin{minipage}[b]{.51\linewidth}
  \centering\includegraphics[width=45mm,height=27mm]{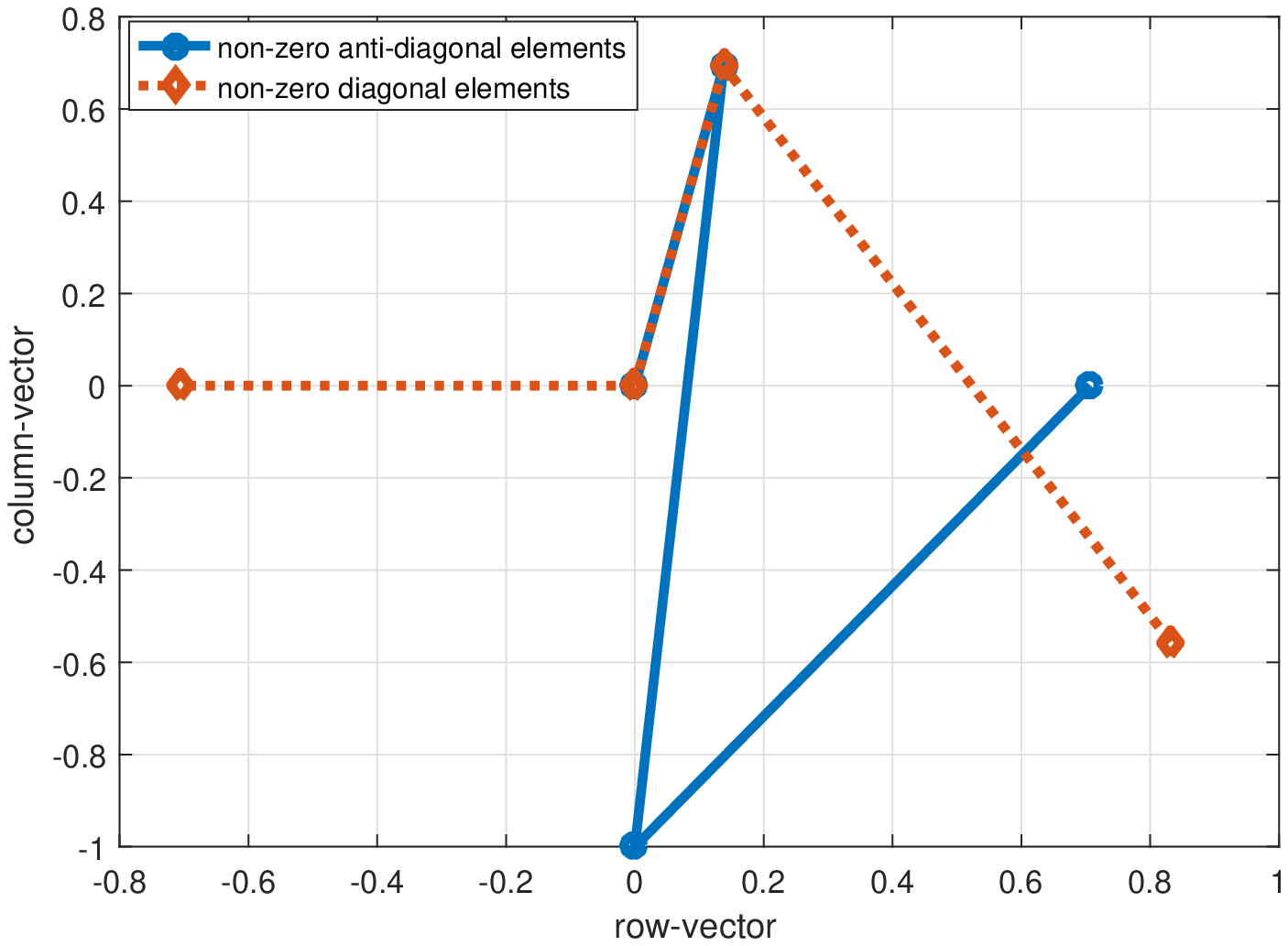}
  \subcaption{}
  \end{minipage}
  \caption{\small Line graphs for different sizes of the DFTC-A codeword. (a) Low-distortion of $2\times2$ DFTC-A codeword in (5) and (b) High-distortion of $4\times2$ DFTC-A codeword from (6). \small Both the red and blue lines represent \textbf{non-zero diagonal} and \textbf{anti-diagonal} elements.}
  \label{Fig1}
  \vspace{1ex}
  \end{figure}
   \raggedbottom
  \section{ Proposed GHC Scheme}
  A Golden section \cite{10,11} with continuous geometric proportion can provide a good OSTBC pattern. Encouraged by this, we first design a GH precoding scheme as follows.\\
  \textit{A. \textbf{GH Precoding Matrices}}: Let $N_{T}=2^{q}$ and $\mathbf{W}_{GH}(2^{q})$ be a $2^{q}\times M$ precoding matrix constructed using $M$ columns of the $2^{q}\times 2^{q}$ recursive GH precoding matrices:
  \begin{eqnarray}
  \mathbf{W}_{GH}(2^{q})=\frac{\theta_{G}}{\sqrt{\xi}} \left[\begin{IEEEeqnarraybox*}[][c]{,c/c/c,}
  \mathbf{W}_{H}(2^{q-1})&\mathbf{W}_{H}(2^{q-1})\\
  \mathbf{W}_{H}(2^{q-1})& ({\theta_{G}^{-1}}-{\theta_{G}})\mathbf{W}_{H}(2^{q-1})
  \end{IEEEeqnarraybox*}\right]
  \end{eqnarray}
  where $2\leq q\epsilon N_{T}$ \cite{12}, $\mathbf{W}_{H}(2^{q-1})$ is the $2^{q-1}\times 2^{q-1}$ Hadamard matrix presented in\cite{13}, $\xi=n\{(1+n)^{q}-(1-n)^{q}\}/2^{q}$, $n$ indicates the root of the geometric number, ${\theta_{G}}$ is the Golden number \cite{14}, which satisfies ${\theta_{G}}{\theta_{G}^{-1}}=1$, $({\theta_{G}^{-1}}-{\theta_{G}})=-1$, and its continuous geometric proportion is ${\theta_{G}}:1:1/{\theta_{G}}$, which indicates that the geometric mean between ${\theta_{G}}$ and $1/{\theta_{G}}$ is unity. Thus, we incorporate (9) in (4) and construct the remaining GH precoding matrices as follows:
  \begin{eqnarray}
  \mathbf{W}_{GH,\textit{i}}(2^{q})=(e^{j\pi\delta_{\textup{min}}(\textbf{W}_{GH}(2^{q}))})^{\textit{i}-1}\mathbf{W}_{GH,1}(2^{q}),\textit{i}=2,..,L
  \end{eqnarray}
  where $e^{(\cdot)}$ denotes the exponential function, $\delta_{\textup{min}}(\textbf{W}_{GH}(2^{q}))$ is the MCD based on (7) and (9), $\mathbf{W}_{GH,1}(2^{q})$ is an $N_{T}\times M$ first GH precoding matrix.  Now, we investigate the real and complex Golden numbers with GH precoding as follows:\\
  \textit{\textbf{Case I} (Real Golden Number)}: We consider that ${\theta_{G_{r}}}$ is the real-valued root of $\mu_{\theta_{G_{r}}}(X)=X^{2}-X-1$; ${\theta_{G_{r}}}=\frac{1+\sqrt{5}}{2}$   is known as the real Golden number \cite{10, 14}, and its algebraic conjugate can be obtained as $\bar{\theta_{G_{r}}}=1-{\theta_{G_{r}}}=\frac{1-\sqrt{5}}{2}$. Let $i=3, q=2,n=\sqrt{5}$, $\xi=5$ and ${\theta_{G_{r}}}=\frac{1+\sqrt{5}}{2}$ in (9) and (10) to obtain the $4\times4$ real GH precoding matrices as
  \begin{eqnarray}
  \mathbf{W}_{GH_{r},3}(4)=(-1)^{2}\mathbf{W}_{GH_{r},1}(4),\textit{i}=3,
  \end{eqnarray}
  where $e^{j\pi\delta_{\textup{min}}(\textbf{W}_{GH_{r}}(2^{q}))}=-1$ and
  \begin{eqnarray}
  \mathbf{W}_{GH_{r},1}(4)=\frac{1+\sqrt{5}}{2\sqrt{5}}\left[\begin{IEEEeqnarraybox*}[][c]{c/c/c,c/c/c}
  1&1&1&1\\
  1&-1&1&-1\\
  1&1&-1&-1\\
  1&-1&-1&1
  \end{IEEEeqnarraybox*}\right]
  \end{eqnarray}\
  \textit{\textbf{Case II} (Complex Golden Number)}: Let ${\theta_{G_{c}}}$ be the complex-valued root of $\mu_{\theta_{G_{c}}}(jX)=X^{2}-jX-1$; ${\theta_{G_{c}}}=\frac{j+\sqrt{3}}{2}$   is known as the complex Golden number, and its algebraic conjugate is obtained as $\bar{\theta_{G_{c}}}=\frac{j-\sqrt{3}}{2}$. Similarly, let $i=3, q=2,n=\sqrt{3}$, $\xi=3$ and ${\theta_{G_{c}}}=\frac{j+\sqrt{3}}{2}$ values in (9) and (10) to obtain the $4\times4$ complex GH precoding matrices as
  \begin{eqnarray}
  \mathbf{W}_{GH_{c},3}(4)=(-j)^{2}\mathbf{W}_{GH_{c},1}(4),\textit{i}=3,
  \end{eqnarray}
  where $je^{j\pi\delta_{\textup{min}}(\textbf{W}_{GH_{c}}(2^{q}))}=-j$ and
  \begin{eqnarray}
  \mathbf{W}_{GH_{c},1}(4)=\frac{j+\sqrt{3}}{2\sqrt{3}} \left[\begin{IEEEeqnarraybox*}[][c]{c/c/c,c/c/c}
  1&1&1&1\\
  1&-1&1&-1\\
  1&1&-j&-j\\
  1&-1&-j&j
  \end{IEEEeqnarraybox*}\right]
  \end{eqnarray}
  Note that (12) and (14) are the exact real and complex GH matrices, respectively. Thus, we can summarize the expression of the designed real and complex GHC generators as
  \begin{eqnarray}
  {\mathcal{F}}_{GH}=\{_{{(-j)^{\textit{i}-1}\textbf{W}_{GH_{c},\textit{i}}(2^{q})}\}, for -\textbf{Case II}}^{\{(-1)^{\textit{i}-1}\textbf{W}_{GH_{r},\textit{i}}(2^{q})\}, {for - \textbf{Case I}}}
  \end{eqnarray}
 Now, we can calculate the MCD and MD values using (15) in (2) to make Table I and II based on (7) and (8) for both real $\mathbf{Case I}$ and complex $\mathbf{Case II}$, respectively.\\
\textit{B. \textbf{Distortion-free codeword}}: When a precoded codeword block or blocks contain zero-diagonal elements, they are distortion-free. Using Section III-A, we consider
 $\sqrt{\left(\begin{IEEEeqnarraybox*}[][c]{c/c/c,c/c/c}
  2.0&0.08\\
  -0.08&2.0
  \end{IEEEeqnarraybox*}\right)
  \left(\begin{IEEEeqnarraybox*}[][c]{c/c/c,c/c/c}
  -0.08&2.0\\
  2.0&0.08
  \end{IEEEeqnarraybox*}\right)}
  =\left(\begin{IEEEeqnarraybox*}[][c]{c/c/c,c/c/c}
  \sqrt{0}&2\\
  2&\sqrt{0}
  \end{IEEEeqnarraybox*}\right)$,
  where diagonal element $\sqrt{0}=0$ represents a zero value of geometric mean and is directly related to achieving a distortion-free codeword which exhibit as an error-free feedback link \cite{15}.\\
  \textit{\textbf{Example 2}}, Let $\textit{i}=3, q=2, N_{T}=4, M=2$, and $n=\sqrt{5}$. Then, by substituting (12) in (6), we obtain the $4\times2$ GHC-A third codeword matrix using $\textit{\textbf{Case I}}$ as
  \begin{eqnarray}
  \left [\mathbf{X}_{3}\right] = \left[\begin{IEEEeqnarraybox*}[][c]{,c/c/c,}
  0.0000+0.0000\textit{i}&1.0233-1.0233\textit{i}\\
  1.0233+1.0233\textit{i}&0.0000+0.0000\textit{i}\\
  0.0000+0.0000\textit{i}&1.0233-1.0233\textit{i}\\
  1.0233+1.0233\textit{i}&0.0000+0.0000\textit{i}
  \end{IEEEeqnarraybox*}\right]
  \end{eqnarray}
  Similarly, by substituting (14) in (6), we obtain the $4\times2$ GHC-A third codeword  matrix using \textbf{Case II} as
  \begin{eqnarray}
  \left [\mathbf{X}_{3}\right] = \left[\begin{IEEEeqnarraybox*}[][c]{,c/c/c,}
  0.0000+0.0000\textit{i}&1.1153-0.2988\textit{i}\\
  0.2988+1.1153\textit{i}&0.0000+0.0000\textit{i}\\
  0.0000+0.0000\textit{i}&1.1153-0.2988\textit{i}\\
  0.2988+1.1153\textit{i}&0.0000+0.0000\textit{i}
  \end{IEEEeqnarraybox*}\right].
  \end{eqnarray}
  Using (16) and (17), we plotted Fig. 2 and , by comparison with Fig. 1, we can see distortion-free codeword curve. Thus, we state $\textbf{\textit{Theorem 1}}$ as follows.\par
  \setlength{\extrarowheight}{1pt}
\begin{table}
\centering
\caption{\small Computation of the MCD}
\begin{tabular}{|*{9}{p{8.52cm}|}}
\hline
\centering
\small MCD, $\sqrt{\delta_{\textup{min}}(\textbf{W})}$\\
 \small when $\textit{i}=2, q=2, N_{T}=4$, and $M=2$
 \end{tabular}\\
 \begin{tabular}{|c|c|c|c|c|}
\hline
\multirow{2}*[-.1ex]{\small DFTC[4, 6]}&\multirow{2}*[-.1ex]{\small DC[9, 10]}&\multirow{2}*[-.1ex]{\small HC[8]} &\multicolumn{2}{|c|}{\small Proposed GHC Scheme}\\\cline{4-5}
 & & & $\small \textbf{Case I}(15)$ & $\small \textbf{Case II}(15)$ \\  \hline
  \small 0.84 & 0.89  & 1.00 & 1.00 & 1.00 \\ \hline
 \end{tabular}\\
\end{table}

\begin{table}
\centering
\caption{\small Computation of the MD}
\begin{tabular}{|*{10}{p{8cm}|}}
\hline
\centering
\small MD, $\sqrt{\delta_{\textup{min}}(\mathcal{C})}$\\
 \small when $\textit{i}=2, q=2, N_{T}=4$, and $M=2$
 \end{tabular}\\
 \begin{tabular}{|c|c|c|c|c|c|}
\hline
\multirow{2}*[-.1ex]{\small{b}}&\multirow{2}*[-.1ex]{\small DFTC[4, 6]}&\multirow{2}*[-.1ex]{\small DC[9, 10] }&\multirow{2}*[-.1ex]{\small HC[8]}&\multicolumn{2}{|c|}{\small Proposed GHC}\\\cline {5-6}
 & & & & $\small \textbf{Case I}$ & $\small \textbf{Case II}$\\  \hline
 \small 4 & 2 & 2  & 2 & 2.89 & 2.52 \\ \hline
\small  6 & 4 & 4  & 4 & 5.78 & 5.05 \\ \hline
 \end{tabular}\\
\end{table}

\textbf{\textit{Theorem 1}}: All forms of $\mathbf{X}$ become distortion-free if the precoding matrix $\mathbf{W}$ is a Hadamard or a Golden-Hadamard.\\
\textit{Proof of \textbf{Theorem 1}}: DFTC-A schemes are not distortion-free due to the exponential function of DFT precoding, which leads to non-zero diagonal elements in the codewords as shown in (5) and (6). Let the exponential function of a DFT precoding be approximately equal to the Golden-Hadamard function:
\begin{eqnarray}
e^{\frac{j2\pi (k-1)(l-1)}{N_{T}}}\approx e^{j\pi}=({\theta_{G}^{-1}}-{\theta_{G}})=-1
\end{eqnarray}
where Euler's formula is related to the Golden section by $e^{j\pi}=({\theta_{G}^{-1}}-{\theta_{G}})$ as stated in \cite{14}. Thus, we incorporate (18) in (10) and obtain the distortion free third codeword using (2)
  \begin{eqnarray}
  \left [\mathbf{X}_{3}\right] &=&\frac{(-1)^{2}\theta_{G}}{\sqrt{\xi}}\left[\begin{IEEEeqnarraybox*}[][c]{,c/c/c}
  0&-s_{21}^{*}+s_{11}^{*}\\
  s_{11}-s_{21}&0\\
  0&-s_{21}^{*}+s_{11}^{*}\\
  s_{11}-s_{21}&0
  \end{IEEEeqnarraybox*}\right]
  \end{eqnarray}
   We can see that (19) is completely distortion-free owing to \textbf{null-diagonal} elements in the codeword block or blocks as explained Section IV-B. Thus, $\textbf{\textit{Theorem 1}}$ has been proven.
  \begin{figure}
  \begin{minipage}[b]{.51\linewidth}
  \centering\includegraphics[width=45mm,height=27mm]{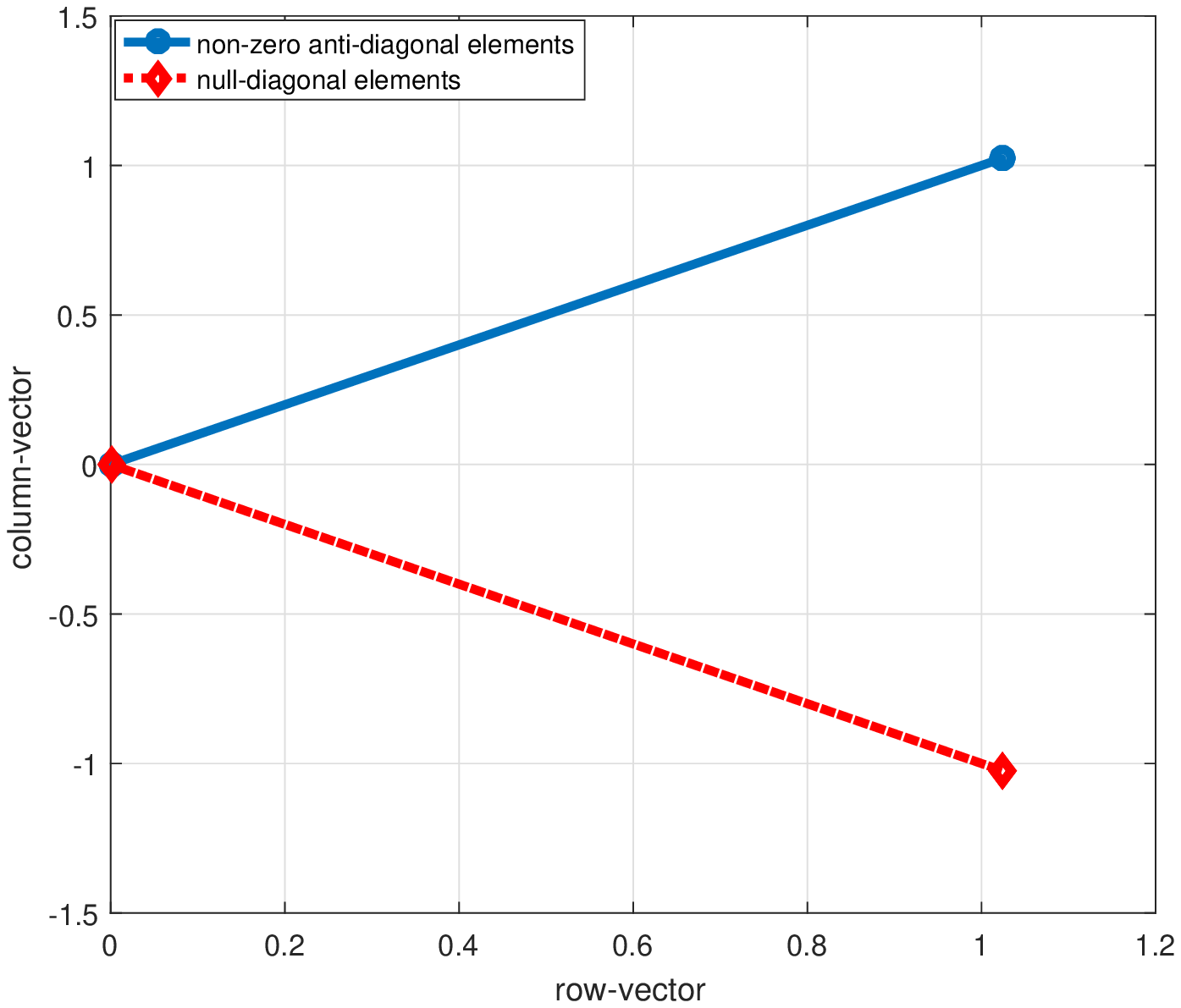}
  \subcaption{}
  \end{minipage}%
  \begin{minipage}[b]{.51\linewidth}
  \centering\includegraphics[width=45mm,height=27mm]{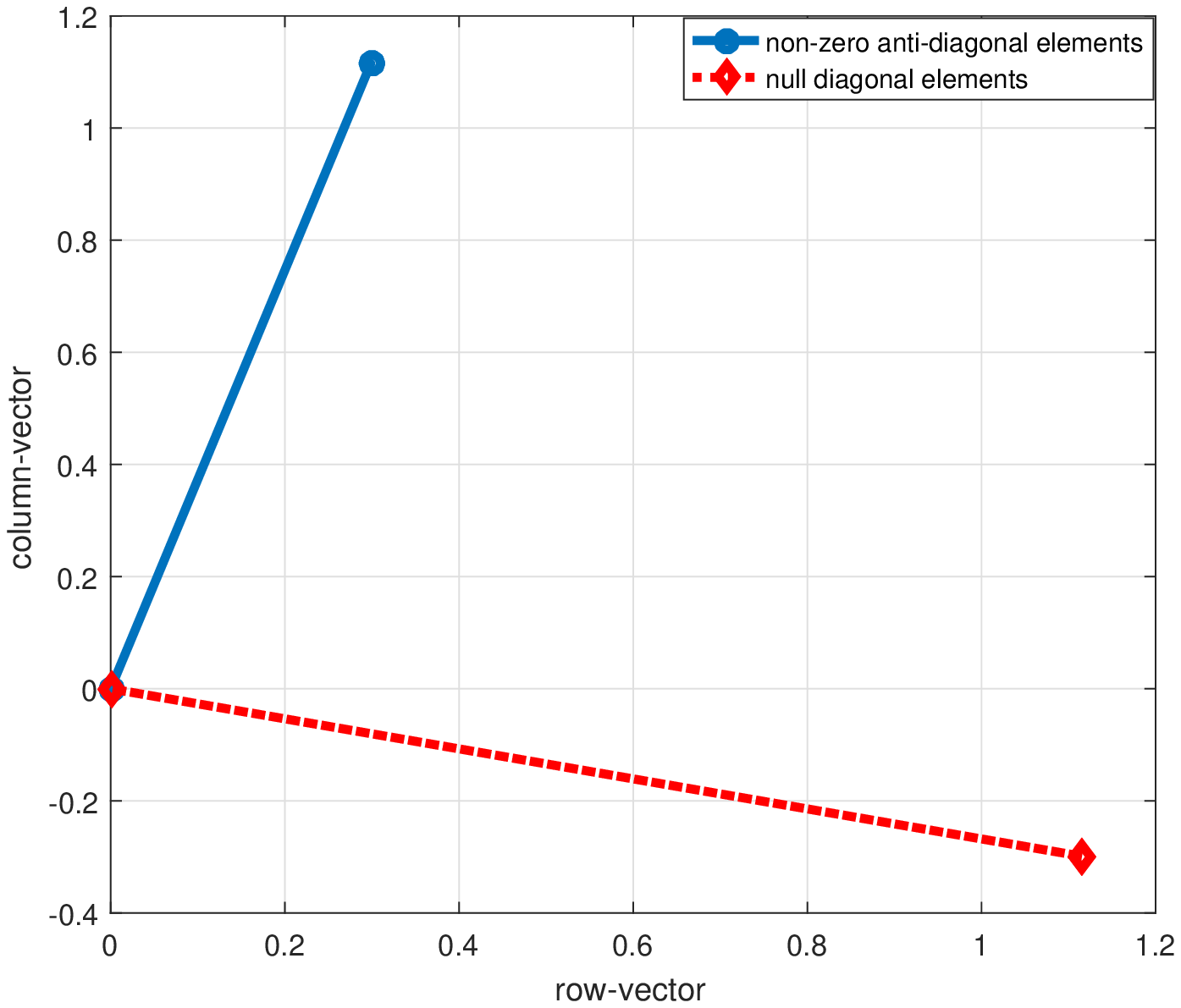}
  \subcaption{}
  \end{minipage}
  \caption{\small Line graphs of the proposed GHC-A codeword: (a) distortion-free $4\times 2$ GHC-A codeword for $\mathbf{Case I}$ obtained using (16), and (b) distortion-free $4\times 2$ GHC-A codeword for $\mathbf{Case II}$ obtained using (17). \small The red lines represent \textbf{null-diagonal} elements and the blue lines represent \textbf{non-zero anti-diagonal} elements.
   \label{Fig2}}
   \vspace{1ex}
  \end{figure}
  \raggedbottom
\section{Pairwise Error Probability (PEP)}
 Let $\mathbf{S}_{\textit{k}}$ and $\mathbf{S}_{\textit{l}}$ be the transmitted and decoded space-time codewords, respectively, and $\textit{l}\neq \textit{k}$. Thus, the union bound on the BER can be written as \cite{16}
 \begin{eqnarray}
 BER\leq \sum_{l\neq k}\frac{e(\mathbf{S}_{k},\mathbf{S}_{l})}{log_{2}2^{b}}\mathbf{P}_{r}(\mathbf{S}_{k}\longrightarrow\mathbf{S}_{l})
 \end{eqnarray}
 where $e(\mathbf{S}_{k},\mathbf{S}_{l})$ is the Hamming  distance between the binary sequences representing $\mathbf{S}_{k}$ and $\mathbf{S}_{l}$, the codeword PEP for an effective channel $\mathbf{\widetilde{h}}$ can be upper bounded by the Chernoff bound \cite{3} as $\mathbf{P}_{r}(\mathbf{S}_{k}\longrightarrow\mathbf{S}_{l}|\mathbf{\widetilde{h}})\leq e^{-{\frac{\gamma_{0}\|\mathbf{\widetilde{h}}\|^{2}}{2\cdot 2^{q}}}}$, $\gamma_{0}$ is the SNR, $\|\mathbf{\widetilde{h}}\|^{2}=\mathbf{h}^{T}\mathbf{W}_{GH}(2^{q})\mathbf{A}\mathbf{W}_{GH}^{H}(2^{q})\mathbf{h}^{*}$, $\mathbf{A}=\bar{\mathbf{S}}\bar{\mathbf{S}}^{H}$ as the covariance distance product matrix, $(\cdot)^{H}$ denotes the Hermitian operator. In the worst case, the average distance, $\mathbf{\overline{A}}=\frac{1}{T}\mathbb{E}[\mathbf{A}]=\frac{1}{T}\sum_{k\neq l}p_{kl}\Xi_{kl}\Xi_{kl}^{H}$ where $p_{kl}$ is the probability of the pair $(\mathbf{S}_{\textit{k}},\mathbf{S}_{\textit{l}})$ and $\Xi=(\mathbf{X}-\mathbf{\widehat{X}})$.\

 If $\mathbf{h}_{max}$ and $\mathbf{h}_{min}$ represent the elements of $\mathbf{h}$ with the largest and smallest magnitude, respectively as explained in \cite{9,10},  then the obtained norm of the effective channel provides the correct, $\|\mathbf{\widetilde{h}}\|_{correct}^{2}=2|{\theta_{G}}|^{2}|\mathbf{h}_{max}|^{2}+(1+|\theta_{G}|^{2}-|\theta_{G}|^{2}|\theta_{G}|^{2})|\mathbf{h}_{min}|^2$ and the incorrect, $\|\mathbf{\widetilde{h}}\|_{incorrect}^{2}=2|{\theta_{G}}|^{2}|\mathbf{h}_{min}|^2+(1+|\theta_{G}|^{2}-|\theta_{G}|^{2}|\theta_{G}|^{2})|\mathbf{h}_{max}|^{2}$ feedback bits, respectively. Thus, the PEP in (20) is bounded by
   \begin{eqnarray}
  \mathbf{P}_{r}(\mathbf{S}_{k}\longrightarrow\mathbf{S}_{l}|\mathbf{\widetilde{h}})\leq e^{-{\frac{1.6180\gamma_{0}(|h_{1}|^{2}+|h_{2}|^{2})}{2\cdot 2^{q}}}}
  \end{eqnarray}
   where $\textit{P}_{c}=1$ is the probability for the correct feedback bit \cite{9,10} and $|\theta_{G}|^{2}=1.6180$ is the optimal choice for \textbf{Case I}. Similarly, we can measure the optimal choice for \textbf{Case II} based on $|\theta_{G_{c}}|^{2}=0.8660+0.50000i$.

\section{Simulation Results}
This section shows some simulation results to demonstrate the proposed GHC schemes for both real $(\textbf{Case I})$ and complex $(\textbf{Case II})$ GH precoding. We implemented the simulations of MISO systems as Monte-Carlo simulations over a block flat Rayleigh fading channel. Moreover, we considered limited feedback cases with an optimal codebook search method at an SNR level of 30 dB. Throughout the simulations, we assumed that $q=2$, $N_{T}=4$, $M=2$, and $L=64$ with six-bit codebooks. We assume $P_{c}=1$ and $|\theta_{G_{r}}|^{2}=1.6180$ $(\textbf{Case I})$ and $|\theta_{G_{c}}|^{2}=0.8660 + 0.5000i$ $(\textbf{Case II})$ for the optimal choice. In a fair comparison, our GHC schemes achieved higher MCD and MD values (calculated as shown in Table I and II) for both 16-QAM and 64-QAM signal constellations and outperformed all prior versions of POSTBC schemes. In Table I, it can be seen that the MCD of the GHC and HC scheme is $\delta_{\textup{min}}(\textbf{W})=1=\sin(90^{0})$, whereas the MCD of the DFTC and DC schemes are $\delta_{\textup{min}}(\textbf{W})=0.84=\sin(57.14^{0})$ and $\delta_{\textup{min}}(\textbf{W})=0.89=\sin(62.87^{0})$, respectively. This means that the degree of freedom was increased by almost $27^0$ to $33^0$ in the GHC and HC scheme compared with that of the DFTC \cite{4, 6} and DC \cite{9, 10} schemes. In contrast, the performance of the HC scheme \cite{8} dramatically worsened than the GHC scheme due to the poorer MD values shown in Table II.\par

In Fig. 3, a $4\times2$ system using a POSTBC with 16-QAM and 64QAM constellations is assumed for the perfect feedback case. For the imperfect feedback with 16QAM shown in Fig. 4, we assumed $\textbf{w}=f(\mathbf{\widetilde{h}})$ where $\mathbf{\widetilde{h}}=\alpha\textbf{h}+\sqrt{1-\alpha^{2}}\textbf{h}_{error}$ and $\alpha=J_{0}(2\pi f_{d}T_{c}\Delta)$, $f_{d}T_{c}=0.01$ is the normalized Doppler frequency and $\Delta=12$ is the feedback delay. Similarly, we expect 64QAM to show a good performance over imperfect feedback. Furthermore, the performance of the DFRSTC \cite{7} is equal to that of the DFTC scheme. The optimal codebook selection method of the GHC scheme provides a 3 to 5 dB array gain compared with all prior versions of POSTBC schemes. Both real \textbf{Case I} and complex \textbf{Case II} demonstrated the better BER performance. The BER performance of the complex GHC scheme is slightly worsened than that of the real GHC scheme because of the reduction of the Golden ratio.
 \begin{figure}[t]	
  \centering
  \includegraphics[width=3in,height=1.5in,keepaspectratio]{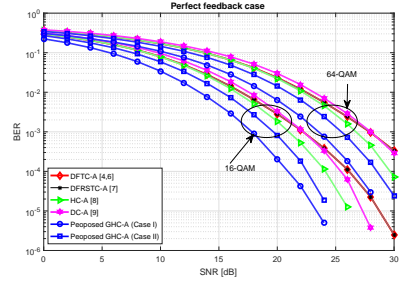}
  \caption{\small BER comparison under perfect feedback. $q=2$, $N_{T}=4$, $M=2$, and $L=64$.}
  \label{Fig3}
  \end{figure}
  \begin{figure}[t]	
   \centering
  \includegraphics[width=3in,height=1.5in,keepaspectratio]{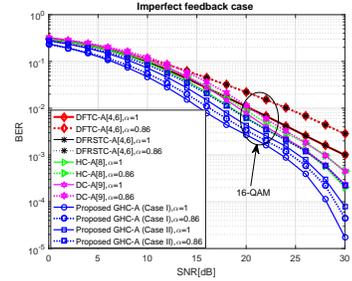}
  \caption{\small BER comparison under imperfect feedback. $q=2$, $N_{T}=4$, $M=2$, $L=64$, $f_{d}T_{c}=0.01$, $\Delta=12$, and $\alpha=0.86$.}
   \label{Fig4}
   \end{figure}
   \raggedbottom
  \section{Conclusions}
 We analyzed codeword distortion along with the MCD and MD problems for all versions of POSTBC schemes. The traditional POSTBC schemes are illustrated worse BER performance by at least $10^{-1}$ steps than the proposed GHC-A scheme due to the smaller MCD and MD values. Moreover, the simulation results confirmed that the proposed GHC scheme outperforms all previous versions of the POSTBC schemes, in terms of BER. Lastly, the GHC scheme can be extended further to be applied in multi-layer wireless networks considering more general cases, which is subjected to future works.
  \ifCLASSOPTIONcaptionsoff
    \newpage
  \fi
 \bibliographystyle{IEEEtran}
  \bibliography{IEEEabrv,Ref}

\end{document}